\documentclass[preprint,showpacs,preprintnumbers,amssymb]{revtex4}
\newcommand{\fslash}[1]{\ooalign{\hfil/\hfil\crcr$#1$}}
\newcommand{\lda}{\stackrel{\leftarrow}{D}}
\begin{document}
\preprint{\vbox{ Submitted to European Physical Journal C 
 }}
\title{Couplings between pion and charmed mesons}
\author{Hungchong Kim}
\email{hung@phya.yonsei.ac.kr}
\author{Su Houng Lee}
\affiliation{%
Institute of Physics and Applied Physics, Yonsei University,
Seoul 120-749, Korea
}
\begin{abstract}
We compute the couplings $DD^*\pi$,$D_1D^*\pi$,$D^*D^*\pi$, $D_1 D_1\pi$
using QCD sum rules.
These couplings are important inputs in the meson exchange model calculations
used to estimate the amount of $J/\psi$ absorption due to pions and rho
mesons in heavy ion collisions.
Our sum rules are constructed at the first order in the pion
momentum $p_\mu$, which give the couplings that are not trivially
related to the soft-pion theorem.
Our calculated couplings, which somewhat depend upon
the values of the heavy meson decay constants,
are $g_{DD^*\pi}=8.2 \pm 0.1$,
$g_{D_1 D^* \pi} = 15.8 \pm 2~{\rm GeV}$,
$g_{D^* D^* \pi}= 0.3 \pm 0.03$ and
$g_{D_1 D_1 \pi}= 0.17 \pm 0.04$.

\end{abstract}

\maketitle

\section{Introduction}
\label{sec:intro}

$J/\psi$ suppression~\cite{Matsui:1986dk} seems to be one of the
most promising signal for QGP formation in RHIC. Indeed the recent
data at CERN~\cite{Abreu:2000ni} show an anomalous suppression of $J/\psi$
formation, which seems to be a consequence of QGP 
formation~\cite{Blaizot:2000ev}. However, before coming to a conclusion,
one has to estimate the amount of $J/\psi$ suppression due to hadronic
final state interactions.
Consequently, there have been a number of works,using various
models~\cite{Kharzeev:1994pz,Kharzeev:1996tw,Martins:1995hd,Wong:2000zb,Wong:2001td,Matinian:1998cb,Haglin:2000xs,Oh:2001qr,Lin:2000ad,Haglin:2001ar},
calculating the $J/\psi$
absorption cross section by light mesons. However, 
the estimate varies by an order of magnitude near threshold.
At this stage, it is necessary to probe each model calculations further to
spell out their corrections and uncertainties.

In the effective meson-exchange model 
approaches~\cite{Matinian:1998cb,Haglin:2000xs,Oh:2001qr,Lin:2000ad,Haglin:2001ar}, important ingredients are
the couplings between open charm mesons and light mesons.
Precise determination of the couplings reduces uncertainties in the
calculation of the dissociation process.
Moreover, a complete set of low-lying open charm mesons 
$D(1870), D^*(2010), D_1(2420)$ 
has to be included.  This is especially necessary in order to probe
the cross section above the threshold.
To provide the basic building blocks for such a model calculation, we
use QCD sum rules~\cite{Shifman:1979bx,Shifman:1979by}
and calculate the couplings $DD^*\pi$,$D^*D^*\pi$,$D_1 D_1\pi$,$D_1D^*\pi$.
These can be used to improve the existing effective model
calculations, which can be applied to future calculation~\cite{Song}
of the $J/\psi$ dissociation process.

At present, there are two approaches in the literature to calculate
the coupling in the QCD sum rule approach,
light-cone QCD sum rules
(LCQSR)~\cite{Belyaev:1995zk,Aliev:1997bp,Colangelo:1998rp} relying
on the operator product expansion near the light cone, and the
conventional QCD sum
rules~\cite{Colangelo:1994es,Colangelo:1995ph,DiBartolomeo:1995ir}
based on the short-distance expansion.
Predictions from LCQSR heavily depend on the twist-2  pion wave function
at the middle
point, whose value, however, has been at the core of debates
for a long time~\cite{Chernyak:1984ej}. Furthermore, the duality
issue in constructing the continuum needs to be carefully
considered~\cite{Kim:2000ku}.
Instead, the conventional QCD sum
rules~\cite{Colangelo:1994es,Colangelo:1995ph,DiBartolomeo:1995ir}
do not suffer from such an uncertainty as the QCD parameters appearing
in this approach are determined from the low-energy theorems such as
PCAC and the soft-pion theorem.
Though  QCD duality again needs to be applied carefully in these
sum rules~\cite{Kim:2000rg,Kim:2000ur},
the uncertainties from the QCD side can be substantially reduced.

In this work, we provide a systematic approach to calculate the
couplings using the conventional QCD sum rules. Our sum rules will
be constructed at the first order in the pion momentum
${\cal O} (p_\mu)$. We improve
the previous sum rule calculations of the $DD^*\pi$ and $D^*D^*\pi$
couplings~\cite{Colangelo:1994es,Colangelo:1995ph,DiBartolomeo:1995ir}.
In particular, QCD duality will be correctly applied according
to the recent suggestion in Refs.~\cite{Kim:2000rg,Kim:2000ur}.
Also in constructing a sum rule for $DD^*\pi$, we advocate
the use of a different structure function.
We then construct similar sum rules for the couplings, $D_1D_1\pi$ and
$D_1 D^* \pi$, which may be important for
future calculation of the $J/\psi$ dissociation process
accompanying $D_1$ meson~\cite{Song}.

\section{OPE for General Correlation function}
\label{sec:general}

In this section, we schematically describe a general procedure
to perform the operator product expansion (OPE) of
the general correlation function with a pion,
\begin{eqnarray}
\Pi (p,q) =
i\int d^4 x e^{iq\cdot x} \langle 0 | T \{ {\bar d}(x) \Gamma_1 c(x),
{\bar c}(0) \Gamma_2 u(0) \} | \pi (p) \rangle \ .
\label{general_cor}
\end{eqnarray}
$\Gamma_1$ and $\Gamma_2$ denote specific gamma matrices
corresponding to the coupling of concern. In later sections,
we will use this general prescription, by simple replacements, to calculate
the OPE for the correlation function of concern.
For instance, to calculate the $D D^* \pi$ coupling,
we will have $\Gamma_1 = \gamma_\mu$ and $\Gamma_2 = i\gamma_5$.

To calculate the coupling that is not trivially related by
chiral symmetry, we will consider the correlation function at the first
order of the pion momentum $p_\mu$ in its expansion. The soft-pion limit
of the correlation function is just a chiral rotation of a vacuum correlation
function ({\it i.e.,without the pion}), which  provides a coupling that is
trivially related to the vacuum correlation function. 
For example, in the case of the pion-nucleon coupling~\cite{Birse:1996zh},
the soft-pion limit leads to Goldberger-Treiman relation with $g_A=1$,
which gives the coupling 30 \% lower than its experimental value.
To determine the coupling more precisely, one needs to go beyond the
soft-pion limit.

\begin{figure}
\caption{The OPE diagrams considered in this work.
The blob in (a) denotes the quark-antiquark component with a pion
and the blob in (b) denotes quark-antiquark-gluon component
with a pion.
}
\label{fig1}

\setlength{\textwidth}{6.1in}   
\setlength{\textheight}{9.in}  
\centerline{%
\vbox to 2.4in{\vss
   \hbox to 3.3in{\includegraphics{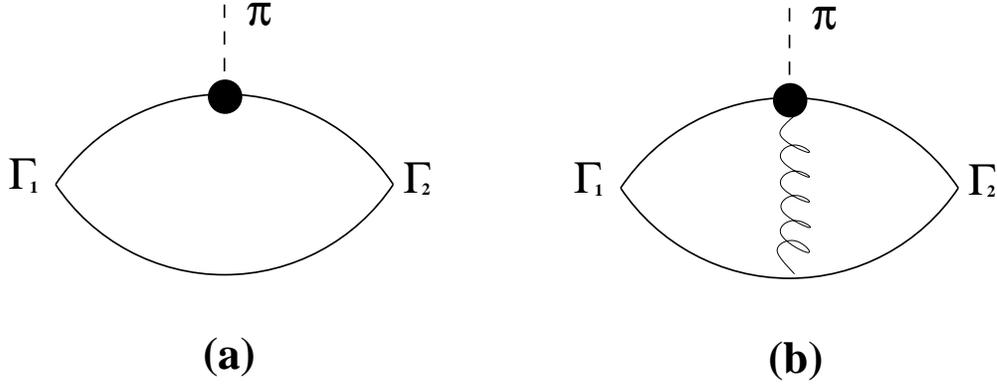}\hss}}
}

\vspace{50pt}
\end{figure}

First, we restrict ourselves to the OPE diagram shown in Fig.~\ref{fig1} (a).
For this diagram, we can rewrite the correlation function
into the form
\begin{eqnarray}
-i \int {d^4 k \over (2\pi)^4}
Tr\left [ i {\fslash{q} -\fslash{k}+m_c \over (q-k)^2 -m^2_c}
\Gamma_2 D_{aa} (k,p) \Gamma_1 \right ] \ .
\label{opetree}
\end{eqnarray}
The Roman indices denote colors.
The free $c$-quark propagator has been used in obtaining this.
$D_{ab} (k,p)$, which is shown by the blob in the figure,
 denotes the quark-antiquark component with a pion.
This can be separated into three pieces depending on the
Dirac matrices involved,
\begin{eqnarray}
 D_{ab} (k,p) = \delta_{ab} \left [ i\gamma_5 A(k,p) +
\gamma_\alpha \gamma_5 B^\alpha (k,p) +
\gamma_5 \sigma_{\alpha \beta} C^{\alpha \beta} (k,p) \right ] \ .
\end{eqnarray}
The three invariant functions of $k,p$ are defined by
\begin{eqnarray}
A(k,p) &=& {1 \over 12} \sum_{n=0}^{\infty} {1\over n !}
\langle 0 | {\bar d} \lda_{\alpha_1}\cdot \cdot \cdot
\lda_{\alpha_n} i\gamma_5 u
| \pi (p) \rangle (2\pi)^4
{ \partial^n \over i\partial k_{\alpha_1} \cdot \cdot \cdot
i\partial k_{\alpha_n} }
\delta^{(4)}(k) \ ,
\nonumber \\
B^\alpha (k,p) &=& {1 \over 12} \sum_{n=0}^{\infty} {1\over n !}
\langle 0 | {\bar d} \lda_{\alpha_1}\cdot \cdot \cdot \lda_{\alpha_n}
\gamma^\alpha \gamma_5 u | \pi (p) \rangle (2\pi)^4
{ \partial^n \over i \partial k_{\alpha_1} \cdot \cdot \cdot i
\partial k_{\alpha_n} }
\delta^{(4)}(k) \ ,
\nonumber \\
C^{\alpha \beta} (k,p)  &=& - {1 \over 24} \sum_{n=0}^{\infty} {1\over n !}
\langle 0 | {\bar d} \lda_{\alpha_1}\cdot \cdot \cdot \lda_{\alpha_n}
\gamma_5 \sigma^{\alpha \beta} u | \pi (p) \rangle (2\pi)^4
{ \partial^n \over i\partial k_{\alpha_1} \cdot \cdot \cdot
i\partial k_{\alpha_n} }
\delta^{(4)}(k) \ .
\end{eqnarray}
Since we are constructing the sum rules at the order ${\cal O} (p_\mu)$,
we need to evaluate the pion matrix elements at ${\cal O} (p_\mu)$.
Noting that the pion matrix elements involved are symmetric
under exchanges of any pair of indices $\alpha_1 \cdot \cdot \cdot
\alpha_n$ and using the soft-pion theorem and PCAC, we straightforwardly 
obtain the followings,
\begin{eqnarray}
\langle 0 | {\bar d} \lda_{\alpha_1}
i\gamma_5 u | \pi (p) \rangle &=& {\langle {\bar q} q \rangle \over f_\pi}
ip_{\alpha_1} \ ,
\nonumber \\
\langle 0 | {\bar d} \lda_{\alpha_1} \lda_{\alpha_2}\lda_{\alpha_3}
i\gamma_5 u | \pi (p) \rangle &=&
{im_0^2 \langle {\bar q} q \rangle \over 12 f_\pi}
(p_{\alpha_1} g_{\alpha_2 \alpha_3} +
p_{\alpha_2} g_{\alpha_1 \alpha_3} + p_{\alpha_3} g_{\alpha_1 \alpha_2}) \ ,
\nonumber \\
\langle 0 | {\bar d}
\gamma^\alpha \gamma_5 u | \pi (p) \rangle &=& ip^\alpha f_\pi \ ,
\nonumber \\
\langle 0 | {\bar d} \lda_{\alpha_1} \lda_{\alpha_2}
\gamma_\alpha \gamma_5 u | \pi (p) \rangle &=&
if_\pi \delta^2 \left [ p_\alpha g_{\alpha_1\alpha_2}{5\over 18} -
(p_{\alpha_2} g_{\alpha\alpha_1}
+p_{\alpha_1} g_{\alpha\alpha_2}){1\over 18} \right ] \ ,
\nonumber \\
\langle 0 | {\bar d} \lda_{\alpha_1}
\gamma_5 \sigma_{\alpha \beta} u | \pi (p) \rangle
&=&
i(p_\alpha g_{\beta\alpha_1}-p_\beta g_{\alpha \alpha_1})
{\langle {\bar q} q \rangle \over 3 f_\pi} \ ,
\nonumber \\
\langle 0 | {\bar d} \lda_{\alpha_1}\lda_{\alpha_2}\lda_{\alpha_3}
\gamma_5 \sigma_{\alpha \beta} u | \pi (p) \rangle \ ,
\nonumber \\
={im_0^2 \langle {\bar q} q \rangle \over 36 f_\pi}
&\Big [& p_\alpha  (g_{\alpha_1 \alpha_2} g_{\alpha_3\beta}
+g_{\alpha_1 \alpha_3} g_{\alpha_2\beta}
+g_{\alpha_3 \alpha_2} g_{\alpha_1\beta} ) - (\alpha \leftrightarrow \beta)
\Big ] \ .
\end{eqnarray}
Here $m_0^2$ and $\delta^2$ are defined via
\begin{eqnarray}
\langle {\bar q} D^2 q \rangle = {m_0^2 \over 2}
\langle {\bar q} q \rangle\ ,
\nonumber \\
\langle 0| {\bar d} g_s {\tilde {\cal G}}^{\alpha\beta}
\gamma_\beta u | \pi(p) \rangle = i \delta^2 f_\pi p^\alpha\ .
\end{eqnarray}
Up to twist-5, these are all the possibilities coming from the
expansion of the quark-antiquark components at the order ${\cal
O}(p_\mu)$.

The additional contribution to the OPE is shown by Fig.~\ref{fig1} (b)
where one gluon emitted from the $c$-quark propagator is combined
with the quark-antiquark component.
Specifically, the $c$-quark propagator with one gluon being attached
is given by~\cite{Reinders:1985sr}
\begin{eqnarray}
-{g_s {\cal G}_{\alpha \beta} \over 2 (k^2 -m^2_c)^2 }
\left [ k^\alpha \gamma^\beta - k^\beta \gamma^\alpha
+ (\fslash{k} + m_c) i \sigma^{\alpha \beta} \right ]\ ,
\end{eqnarray}
where ${\cal G}_{\alpha \beta} = t^A G_{\alpha \beta}$.
The color matrices $t^A$ are normalized via $Tr(t^A t^B) = \delta^{AB}/2$.
Taking the gluon stress tensor into the quark-antiquark component, one
can write down the correlation function into the form
\begin{eqnarray}
\Pi(p,q) &=& 2i \int {d^4 k \over (2\pi)^4}
Tr \Bigg \{ {2(q-k)_\theta \gamma_\delta +(\fslash{q} -\fslash{k}+im_c)
i\sigma_{\theta\delta} \over \left [ (q-k)^2-m_c^2 \right ]^2 }
\nonumber \\
&&\times \Gamma_2
\left [ \gamma_5 \sigma_{\rho\lambda} B^{\rho\lambda\theta\delta}(k,p)
+\gamma^\tau \epsilon^{\theta\delta\alpha\beta}
D_{\tau\alpha\beta}(k,p) \right ]
\Gamma_1
 \Bigg \} \ .
\label{opeglu}
\end{eqnarray}
At the order ${\cal O} (p_\mu)$, the two functions appearing
above are given by
\begin{eqnarray}
B^{\rho\lambda\theta\delta} &=& -{m_0^2 \langle {\bar q} q \rangle
\over 12\times 32 f_\pi} (g^{\rho\theta}g^{\lambda\delta}
-g^{\rho\delta}g^{\lambda\theta})p^\alpha
(2\pi)^4 {\partial\over\partial k_\alpha}
\delta^{(4)}(k)\ ,
\\
D_{\tau\alpha\beta} &=& -{i \delta^2 f_\pi \over 3 \times 32}
(p_\alpha g_{\tau\beta} - p_\beta g_{\tau \lambda})
(2\pi)^4 \delta^{(4)}(k) \ .
\end{eqnarray}
Another function involving the pion matrix element of the form
\begin{eqnarray}
\langle 0| {\bar d} \gamma_5 \gamma_\tau g_s {\cal G}_{\theta\delta} u|\pi(p)
\rangle
\end{eqnarray}
is of the order ${\cal O} (m_q)$, which therefore has been neglected.

Once the Dirac matrices $\Gamma_i~(i=1,2)$ are given, we can 
straightforwardly calculate the corresponding
OPE from Eqs.~(\ref{opetree}) and (\ref{opeglu}).
These give the OPE up to twist-5 at the order ${\cal O} (p_\mu)$.
Below, we will use this formalism to calculate the
couplings $DD^*\pi$, $D^*D^*\pi$,$D_1D_1\pi$, and $D^*D_1\pi$
by choosing appropriate Dirac matrices for them.

\section{Sum rule for $D D^* \pi$}
\label{sec:ddspi}

In this section, we construct a sum rule for the $D D^* \pi$ coupling
using the correlation function
\begin{eqnarray}
\Pi_\mu (p,q) =
i\int d^4 x e^{iq\cdot x} \langle 0 | T \{ {\bar d}(x) \gamma_\mu c(x),
{\bar c}(0) i\gamma_5 u(0) \} | \pi (p) \rangle\ .
\end{eqnarray}
As two momenta are involved in the correlation function,
one can separate the correlation function into the following two pieces
\begin{eqnarray}
\Pi_\mu = F_1 (p,q) p_\mu + F_2 (q,p) q_\mu \ .
\label{sep1}
\end{eqnarray}
We construct a sum rule for $F_1 (p=0,q)$, which gives
a coupling determined at the order ${\cal O}(p_\mu)$.
Defining the $D^*D\pi$ coupling by~\cite{Belyaev:1995zk}
\begin{eqnarray}
\langle D^*(q)|D(q-p)\pi(p)\rangle = g_{D^*D\pi} p\cdot \epsilon
\end{eqnarray}
and using
\begin{eqnarray}
\langle D | {\bar c} i\gamma_5 u |0\rangle = {m_D^2 f_D \over m_c}
\;; \quad
\langle 0 | {\bar d} \gamma_\mu c |D^*\rangle = m_{D^*} f_{D^*} \epsilon_\mu\ ,
\end{eqnarray}
the low-lying pole contribution to $F_1 (p=0,q)$
is given by
\begin{eqnarray}
-{m_D^2 m_{D^*} f_D f_{D^*} g_{DD^*\pi}
\over
             m_c (q^2-m^2_{D^*}) (q^2 - m_D^2)}\ .
\label{phen1}
\end{eqnarray}

A slightly different sum rule using the same correlation function can be
found in
Refs.~\cite{Colangelo:1994es,Colangelo:1995ph}.
Specifically, the correlation function in that approach was decomposed into 
\begin{eqnarray}
\Pi_\mu = A p_\mu + B (2q-p)_\mu
\label{col}
\end{eqnarray}
and a sum rule was constructed for the function $A$. Note, by
comparing this with Eq.~(\ref{sep1}), one can immediately see that
$A=F_1+F_2/2$. Thus, in the expansion in terms of the external
momentum $p_\mu$, the function $A$ involves the term at the zeroth
order in $p_\mu$ ({\it i.e.} $F_2$), which can be trivially
obtained from a vacuum correlation function via the soft-pion theorem,
as well as the term of the first
order in $p_\mu$ ({\it i.e.,} $F_1$). To avoid the trivial contribution
obtained from the soft-pion theorem, we choose to work with the
decomposition of Eq.~(\ref{sep1}).
Furthermore, as
mentioned in Ref.~\cite{Belyaev:1995zk}, the function $A$ can
contain some contribution from a possible resonance (scalar
particle $D_0$), which can give an additional uncertainty in the
prediction.

Following the general strategy given in Sec.~\ref{sec:general},
we obtain the OPE for the correlation function,
\begin{eqnarray}
F_1(q,p=0) &=& {1 \over q^2-m_c^2}
\left [ m_c f_\pi -{2 \over 3} {\langle {\bar q} q\rangle \over f_\pi }
\left ( 2-{m_c^2 \over q^2-m_c^2} \right )
-{10 \over 9} {\delta^2 f_\pi m_c \over q^2 -m_c^2 }
\left (1+ {m_c^2 \over q^2-m_c^2 } \right )
\right ]
\nonumber \\
&+&{m_0^2\over 6 f_\pi} \langle{\bar q} q\rangle
\left [ {5 \over 6(q^2-m_c^2)^2} + {4 m_c^2 \over 3(q^2-m_c^2)^3}
-{4 m_c^4 \over (q^2-m_c^2)^4}\right ]\ .
\label{ope1}
\end{eqnarray}
Note, the leading OPE has a simple pole in $q^2$. According to QCD
duality, higher resonance contributions lying along the positive
$q^2$ are matched with the imaginary part of the OPE above a
certain threshold $S_0$ which is taken much higher than the
low-lying pole. Since the simple pole structure $1/(q^2-m_c^2)$
does not have nonanalytic structures in the duality region ($q^2
\ge S_0$), it should not pick up the continuum contribution.
However, it is an often practice that the double-variable
dispersion relation is used to obtain a spectral density for a
given OPE. Then by naively restricting the dispersion integral
below the continuum threshold, one picks up the continuum
contribution even from the OPE of the form $1/(q^2-m_c^2)$. In
this prescription, the continuum contribution is a simple (and
unphysical) pole at the continuum
threshold~\cite{Kim:2000rg,Kim:2000ur,Kim:2000ku}. (See for
example Eq.(3.12) of Ref.~\cite{DiBartolomeo:1995ir}.) This pole
at the continuum threshold does not resemble at all the higher
resonance contributions lying along the positive $q^2$.
In fact, this pole at the continuum is mathematically spurious.
To illustrate this in detail, let's determine the spectral density for
the OPE $1/(q^2-m_c^2)$ from the double dispersion relation when 
the external momentum is zero,
\begin{eqnarray}
{1 \over q^2 - m_c^2} = \int_0^\infty ds {b(s) \over (s-q^2)^2}\ .
\label{double}
\end{eqnarray}
Under the successive Borel transformations~\cite{Kim:2000ur}, one can 
determine the spectral function 
\begin{eqnarray}
b(s) = - \theta (s-m_c^2)\ .
\label{bs}
\end{eqnarray}
When we put it back to the double dispersion relation, we have
to reproduce the OPE $1/(q^2-m_c^2)$. Anything additional
to it is mathematically spurious. Using Eq.(\ref{bs}) in
Eq.(\ref{double}) and doing the integration by part, we obtain
\begin{eqnarray}
\int_0^\infty ds {- \theta (s-m_c^2) \over (s-q^2)^2}= 
{1 \over q^2 - m_c^2} + 
{\theta (s-m_c^2) \over s-q^2} \Bigg |^\infty_0
\end{eqnarray} 
The second term is mathematically spurious as it is additional to the
one that we had started with.  But when it is restricted by the continuum
threshold $S_0$, the upper limit is changed to $S_0$
and the second term has a pole at the continuum threshold. 
However,
since the pole comes from the spurious term, its contribution to the
sum rule is spurious.

Nonetheless, one may argue from an intuition that the continuum
contribution should be present as the current can couple to higher
resonances. In fact, it may be possible to build such a
contribution if one uses a more sophisticated current than the
simple current of the form ${\bar q} \Gamma c$. We believe that
the absence of the continuum is due to a limitation of the
current of the form ${\bar q} \Gamma c$. But we believe that,
as we demonstrated above, it is {\it ad hoc} 
to build the continuum contribution from the OPE of the form
$1/(q^2-m_c^2)$.

We now match Eq.(\ref{phen1}) with Eq.(\ref{ope1}) to get a sum rule
for the $D^* D \pi$ coupling.  For simplicity, we neglect the mass
difference between $D^*$ and $D$ and set them to $m_{D^*} =
m_D =m_{av}\equiv (m_{D^*}+m_D)/2$.
Under the Borel transformation (with the Borel mass $M^2$), the final
sum rule reads,
\begin{eqnarray}
g_{D D^* \pi} f_D f_{D^*}  +
T_1 M^2 &=& {m_c \over m_{av}^3} M^2 e^{(m^2_{av} - m_c^2)/M^2} \Bigg \{
m_c f_\pi -{4\over 3 f_\pi} \langle{\bar q}q\rangle
\nonumber \\
&+&\left [ -{2\over 3f_\pi} m_c^2 \langle{\bar q}q\rangle +
{10 \over 9}\delta^2f_\pi m_c -{5\over 36 f_\pi} m_0^2 \langle{\bar q}q\rangle
\right ] {1\over M^2}
\nonumber \\
&+&\left [-{5\over 9} \delta^2 f_\pi m_c^3 +{1\over 9f_\pi} m_0^2
m_c^2 \langle{\bar q}q\rangle \right ] {1\over M^4}
+ {m_0^2 m_c^4 \langle{\bar q}q\rangle \over 9f_\pi M^6}
\Bigg \} \ .
\label{bddspi}
\end{eqnarray}
Here $T_1$ denotes the transitions from the low-lying resonance to
higher resonances.  We will linearly fit the RHS within a Borel window
to determine the coupling as well as the transition strength $T_1$.

\section{Sum rule for $D^* D^* \pi$}
\label{sec:dsdspi}

We now construct a sum rule for the $D^* D^* \pi$ coupling.
The $D^* D^* \pi$ sum rule can be constructed from the correlation function
(by setting $\Gamma_1 = \gamma_\mu$ and $\Gamma_2 = \gamma_\nu$ in
Eq.~(\ref{general_cor})).
\begin{eqnarray}
\Pi_{\mu\nu} (p,q) =
i\int d^4 x e^{iq\cdot x} \langle 0 | T \{ {\bar d}(x) \gamma_\mu c(x),
{\bar c}(0) \gamma_\nu u(0) \} | \pi (p) \rangle\ .
\label{dsdspicor}
\end{eqnarray}
Saturating the correlation function by the $D^*$ intermediate
state and introducing the coupling via
\begin{eqnarray}
\langle D^*(q,\epsilon_2) | \pi(p) D^*(q-p,\epsilon_1) \rangle
=i {2\over f_\pi} g_{D^*D^*\pi}\epsilon_{\alpha\beta\mu\nu}
\epsilon_1^\alpha \epsilon_2^\beta p^\mu q^\nu\ ,
\label{dsdsphen}
\end{eqnarray}
the low-lying pole contribution at the first order in $p_\mu$
is given by~\cite{DiBartolomeo:1995ir}
\begin{eqnarray}
- {2 g_{D^*D^*\pi} f^2_{D^*} m^2_{D^*}
\over f_\pi (q^2-m_{D^*}^2)^2} \epsilon_{\alpha\beta\mu\nu} p^\mu q^\nu\ .
\end{eqnarray}

The OPE part can be computed by following the general
prescription described in Sec.~\ref{sec:general}.
After taking out the common factor of
$\epsilon_{\alpha\beta\mu\nu} p^\mu q^\nu$, we obtain the OPE side
\begin{eqnarray}
{f_\pi \over q^2-m_c^2}
+ {2\over 3f_\pi} {m_c \langle {\bar q} q \rangle \over (q^2-m_c^2)^2}
+ {8\over 9} {f_\pi \delta^2 \over (q^2-m_c^2)^2}
-{10\over 9} {f_\pi \delta^2 m_c^2 \over (q^2-m_c^2)^3}
-{2\over 3f_\pi} {m_c^3 m_0^2 \langle {\bar q} q \rangle \over (q^2 -m_c^2)^4}
\ .
\label{sspope}
\end{eqnarray}
The terms involving $m_0^2$ are different from the one in
Ref.~\cite{DiBartolomeo:1995ir}.
By matching the two sides, we obtain
the sum rule for the $D^*D^*\pi$ coupling
\begin{eqnarray}
g_{D^*D^*\pi} f^2_{D^*}
&+& T_2 M^2
=  {f_\pi \over 2 m_{D^*}^2} e^{(m_{D^*}^2-m_c^2)/M^2}
\nonumber \\
&\times& \Bigg [ f_\pi M^2
- {2\over 3 f_\pi} m_c \langle{\bar q}q \rangle -
{8\over 9} f_\pi \delta^2
-{5\over 9} f_\pi \delta^2 {m_c^2 \over M^2}
+ {m_c^3 m_0^2 \langle{\bar q}q \rangle \over 9 f_\pi M^4}
\Bigg ]\ .
\label{dsdspisum}
\end{eqnarray}
Here, $T_2$ denotes the transitions from $D^* \rightarrow
higher~ resonance ~states$.

\section{Sum rule for $D_1 D_1 \pi$}
\label{sec:d1d1pi}

For the coupling $D_1 D_1 \pi$, we use the correlation function
involving axial-vector currents,
\begin{eqnarray}
i\int d^4 x e^{iq\cdot x} \langle 0 |
T \{ {\bar d}(x) \gamma_\mu \gamma_5 c(x),
{\bar c}(0) \gamma_\nu \gamma_5 u(0) \} | \pi (p) \rangle \ .
\end{eqnarray}
Comparing this correlation function with Eq.~(\ref{dsdspicor}),
one can easily see that the OPE for this correlation function
can be obtained by replacing $m_c \rightarrow -m_c$ in
Eq.~(\ref{sspope}).
We introduce the $D_1D_1\pi$ coupling similarly as Eq.~(\ref{dsdsphen}).
Then it is a simple matter to construct a sum rule for the $D_1 D_1 \pi$
coupling.  Namely, by replacing $m_c \rightarrow -m_c$ in
Eq.~(\ref{dsdspisum}), we have
\begin{eqnarray}
g_{D_1D_1\pi} f^2_{D_1}
&+& T_3 M^2
= {f_\pi \over 2 m^2_{D_1}} e^{(m_{D_1}^2-m_c^2)/M^2}
\nonumber \\
&\times& \Bigg [ f_\pi M^2
+ {2\over 3 f_\pi} m_c \langle{\bar q}q \rangle -
{8\over 9} f_\pi \delta^2
-{5\over 9} f_\pi \delta^2 {m_c^2 \over M^2}
- {m_c^3 m_0^2 \langle{\bar q}q \rangle \over 9 f_\pi M^4}
\Bigg ]\ .
\label{11pisum}
\end{eqnarray}
Again, $T_3$ denotes the transitions from $D_1 \rightarrow
higher~ resonance~ states$.

\section{Sum rule for $D_1 D^* \pi$}
\label{sec:d1dspi}

The $D_1 D^* \pi$ coupling can be calculated from the correlation
function
\begin{eqnarray}
i\int d^4 x e^{iq\cdot x} \langle 0 |
T \{ {\bar d}(x) \gamma_\mu c(x),
{\bar c}(0) \gamma_\nu \gamma_5 u(0) \} | \pi (p) \rangle \ .
\end{eqnarray}
Refs.~\cite{Aliev:1997bp,Colangelo:1998rp} calculated the coupling
using the light-cone QCD sum rules.
Both references considered the structure
function proportional to $g_{\mu\nu}$ but the OPE in either approaches
are different. 
Here we choose to work with the structure function proportional
to $q_\mu p_\nu + q_\nu p_\mu$, which turns out to be rather simple.


In constructing the phenomenological side, we follow
Ref.~\cite{Aliev:1997bp} where the structure
proportional to $q_\mu p_\nu + q_\nu p_\mu$
is given by
\begin{eqnarray}
g_{D_1 D^* \pi} {m_{D_1} \over m_{D^*}} {f_{D^*} \over q^2 - m_{D^*}^2}
{f_{D_1} \over q^2 - m_{D_1}^2}\ .
\end{eqnarray}
Note, $g_{D_1 D^* \pi}$ has one dimension due to the
way that the coupling is introduced in Ref.~\cite{Aliev:1997bp}, which is 
in contrast to the other dimensionless couplings.
The OPE side can be calculated straightforwardly.
It takes the simple form of
\begin{eqnarray}
-f_\pi \left [ {1\over q^2-m_c^2} -
{10 \over 9} {m_c^2 \over (q^2 -m_c^2)^3}
\right ] \ .
\end{eqnarray}
The terms containing $\delta^2$ from
the expansion of $A(k,p)$ is canceled
with the similar term coming from $D_{\tau \alpha \beta}$.
Setting $m_{D^*} = m_{D_1}=  {\bar m} \equiv (m_{D^*}+m_{D_1})/2 $
and matching the two sides,
we obtain
\begin{eqnarray}
g_{D_1 D^* \pi} f_{D^*}f_{D_1} + T_4 M^2
= f_\pi e^{({\bar m}^2 -m_c^2)/M^2}
\left [ M^2 -{5\over 9} {m_c^2 \over M^2} \right ] \ .
\label{bd1dspi}
\end{eqnarray}

\section{Analysis and results}
\label{sec:analysis}

\begin{figure}
\caption{The Borel curves for the $DD^*\pi$ coupling
from  Eqs.~(\ref{bddspi}).
Here the RHS of the equation is plotted with respect to
the Borel mass $M^2$.
}
\label{fig2}

\setlength{\textwidth}{6.1in}   
\setlength{\textheight}{9.in}  
\centerline{%
\vbox to 2.4in{\vss
   \hbox to 3.3in{\includegraphics{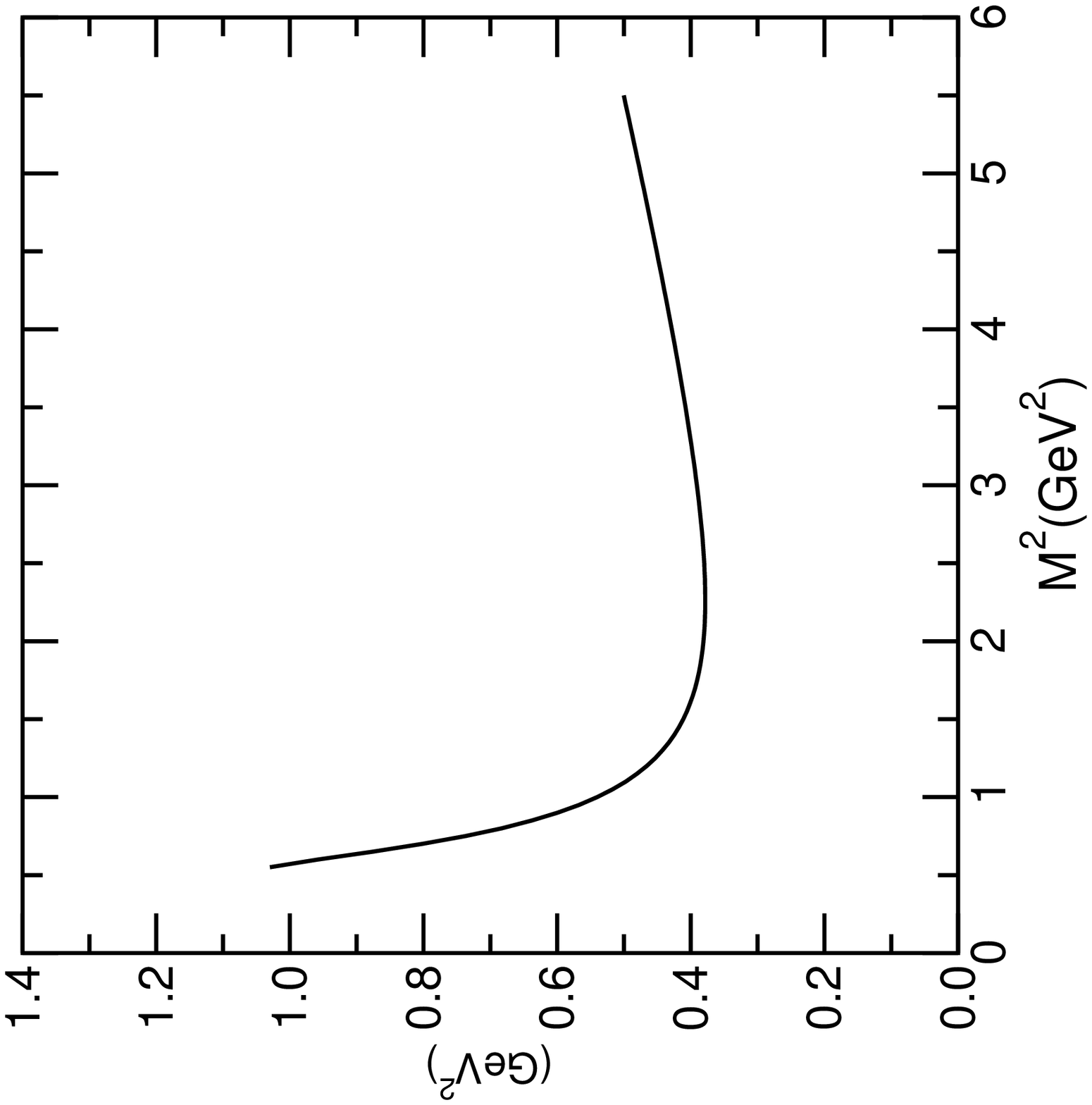}\hss}}
}

\vspace{100pt}
\end{figure}

\begin{figure}
\caption{The Borel curves for the $D_1D^*\pi$ coupling
from  Eq.(\ref{bd1dspi}).
Here the RHS of the equation is plotted with respect to
the Borel mass $M^2$.
}
\label{fig3}

\setlength{\textwidth}{6.1in}   
\setlength{\textheight}{9.in}  
\centerline{%
\vbox to 2.4in{\vss
   \hbox to 3.3in{\includegraphics{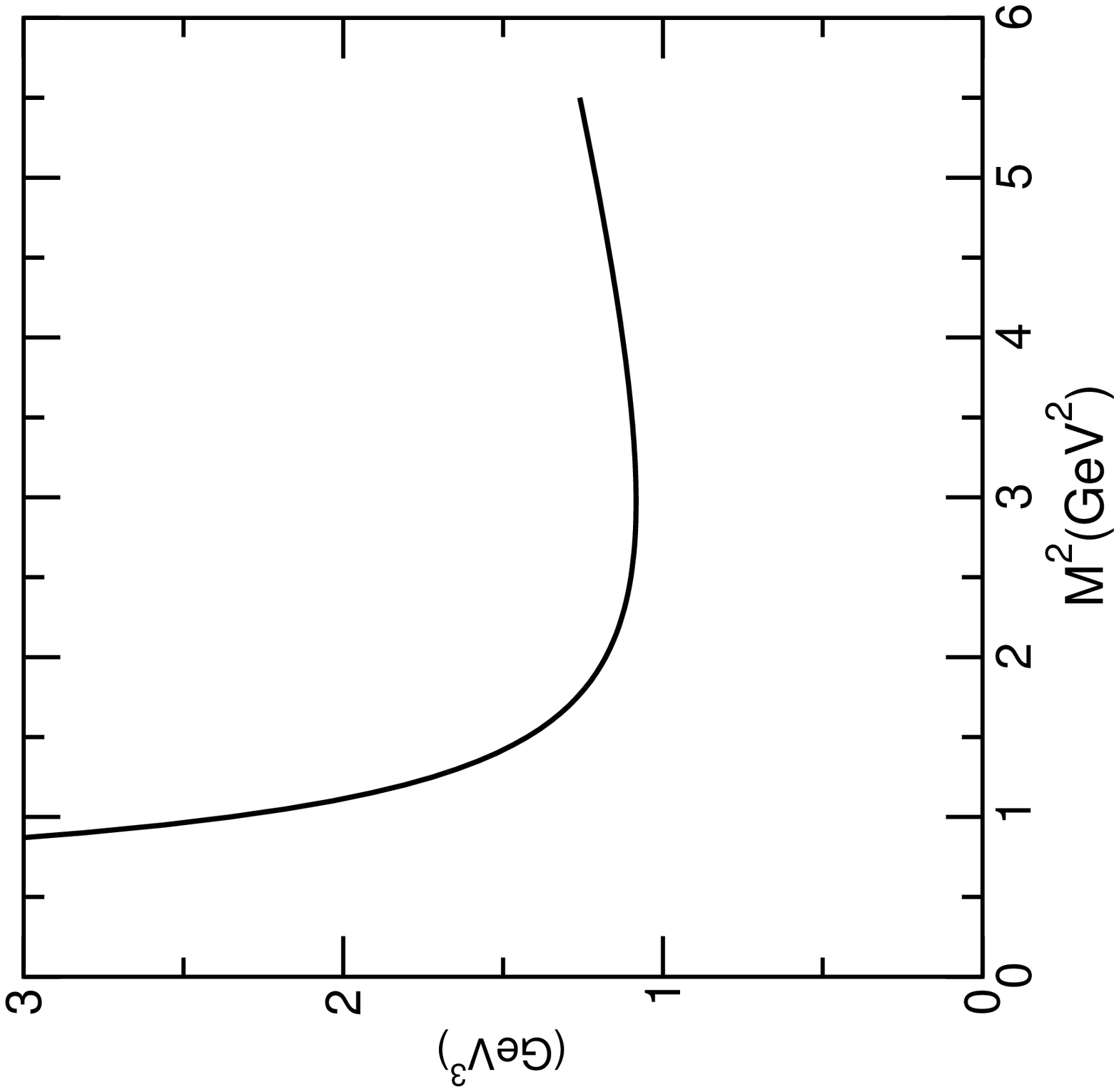}\hss}}
}

\vspace{100pt}
\end{figure}

In our analysis, we use the following set of the QCD parameters,
\begin{eqnarray}
&&m_0^2 = 0.8~{\rm GeV}^2
\;; \quad
\langle {\bar q} q \rangle = (-0.24~{\rm GeV})^3
\;; \quad
\delta^2 =0.2~{\rm GeV}^2
\nonumber \\
&&m_c =1.34 ~{\rm GeV}
\;; \quad
f_\pi =131~{\rm MeV}\ .
\end{eqnarray}
For the hadron masses, we use~\cite{Groom:2000in}
\begin{eqnarray}
m_{D}=1.87~{\rm GeV}
\;; \quad
m_{D^*}=2.01~{\rm GeV}
\;; \quad
m_{D_1}=2.42~{\rm GeV}\ .
\end{eqnarray}
We plot the Borel curves for the couplings $DD^*\pi$, $D_1D^*\pi$
in Figs.~\ref{fig2} and \ref{fig3} using 
Eqs.~(\ref{bddspi}) and (\ref{bd1dspi}) respectively.
The corresponding
curves for the $D^*D^*\pi$ and $D_1D_1\pi$ couplings are shown in
Fig.~\ref{fig3}. To get the couplings, we need to fit each curve
with a straight line within an appropriately chosen Borel window.
The intersection between the best fitting curve and the vertical
axis at $M^2=0$ gives the values $f_D f_{D^*} g_{DD^*\pi}$,
$f_{D_1}f_{D^*} g_{D_1D^*\pi}$, $f^2_{D^*}g_{D^*D^*\pi}$ and
$f^2_{D_1}g_{D_1D_1\pi}$. All the curves are well-fitted with a
straight line above minimum Borel mass, which depending on sum rules,
ranging from 2 - 3 GeV$^2$.  In that region, 
the higher dimensional terms are suppressed. 
When we shift the Borel window by 0.5 GeV$^2$, the results reduce by 10 \%.

Table~\ref{tab:table1} shows the best fit values and the chosen
Borel window. For $f_D f_{D^*} g_{DD^*\pi}$, our value is
substantially smaller than 0.51 GeV$^2$ obtained from the light-cone QCD sum
rule analysis~\cite{Belyaev:1995zk}. The origin of
the difference may be traced back to the use of the asymptotic pion wave
functions. In the light-cone sum rules, the values of the
 twist-2 and twist-3
wave functions at the middle point, $\varphi_\pi(1/2) \sim1.2$ and
$\varphi_p(1/2)\sim 1.5$, enter as leading terms
in the OPE. But in our sum rules, these middle points are replaced
by the integrated strengths of the wave functions $\int_0^1 du
\varphi_\pi(u) =1$, $\int_0^1 du \varphi_p(u) =1$, which are
well-fixed by low-energy theorems. Though our approach is
different from the similar calculations in
Refs.~\cite{Colangelo:1994es,DiBartolomeo:1995ir}, our result for
$f_D f_{D^*} g_{DD^*\pi}$ agrees with them.  Also our value of
$f_{D^*}^2 g_{D^*D^*\pi}$ agrees with the results in
Ref.~\cite{Colangelo:1998rp}.
Our result for $f_{D_1} f_{D^*} g_{D_1D^*\pi}$, 0.948 GeV$^3$, is 
somewhat larger than 0.68 GeV$^3$ from Ref.\cite{Aliev:1997bp},
which is obtained from the structure function proportional to $g_{\mu\nu}$.

To get the couplings, we need to determine $f_D$, $f_{D_1}$ and $f_{D^*}$.
One may calculate these using QCD sum rules of two-point correlation
function in the vacuum. Currently
these values are not known precisely.
According to Ref.~\cite{Colangelo:1998rp}, they are 
$f_D=170$ MeV, $f_{D^*}=220$ MeV and $f_{D_1}=240$ MeV. A somewhat
different set of the decay constants can be found in Ref.\cite{Aliev:1997bp},
$f_D=160$ MeV, $f_{D^*}=240$ MeV and $f_{D_1}=300$ MeV.
Using these, we obtain the coupling constants,
\begin{eqnarray}
&&g_{DD^*\pi}=8.29~ (8.07)
\;; \quad
g_{D_1 D^* \pi} = 17.95 ~(13.6)~{\rm GeV} \nonumber \\
&& g_{D^* D^* \pi}= 0.33~(0.278)
\;; \quad
g_{D_1 D_1 \pi}= 0.208~(0.133)\ .
\end{eqnarray}
where the numbers (the ones in the 
parenthesis) are obtained by using the decay constants
given in Ref.~\cite{Colangelo:1998rp} (Ref.\cite{Aliev:1997bp}).
To make a more precise prediction, one may need to determine
the decay constants precisely.
To summarize, we present in table~\ref{tab:table2} our 
results in comparison
with the other previous calculations.

These couplings are important ingredients for estimating the
absorption cross section of $J/\psi$ by $\pi$ mesons.
Up to now meson exchange models in the calculation of the absorption
cross section
are based on a effective chiral lagrangian with only pseudoscalars
($D, \pi$) and vector mesons ($J/\psi, D^*$).  We believe that the addition
of the axial partner of $D^*$, namely the $D_1$ meson, will reduce the
value for the existing calculation.   The reason is the following.
Consider the dissociation of $J/\psi$ by pions into $D$ and $D^*$.
In the existing meson exchange calculations, the diagrams contributing
to the process are the contributions from the s-channel $D$ meson, the
t-channel $D^*$ mesons and the direct four point coupling, which gives
a non-trivial contribution.  The form of the couplings are obtained
from the chiral SU(4) lagrangian with vector mesons introduced in
the massive Yang-Mills approach.
We believe that the addition of $D_1$ meson
will partly cancel the contribution from the direct four point coupling.
This is so because this is precisely how the Adler consistency condition
for the $\rho -\pi$ forward scattering amplitude
is obtained in the massive Yang Mills approach~\cite{Lee:1995wx}.  
Namely, one has to
introduce the axial partner of $\rho$, namely $a_1$ meson, whose contribution
in the s-channel will cancel the direct four point coupling of the
$\rho \rho \pi \pi$ and make the amplitude vanish in the soft pion limit.
Similar cancellation will occur between the direct four point coupling
of $J/\psi-\pi-D-D^*$ and the 
$J/\psi+\pi \rightarrow D_1 \rightarrow D+D^*$
contribution.  Of course, how big the cancellation actually is in this
particular case has to be studied in detail and that will be done in
the future work  reported in Ref.~\cite{Song}.

\acknowledgments 
The work by Hungchong Kim was supported by the Korea Research Foundation Grant
KRF-2001-015-DP0104. The work by Su Houng Lee was supported by the KOSEF 
1999-2-111-005-5, by the Yonsei University Research Grant, and by
the Ministry of Education 2000-2-0689.

\eject

\begin{table}
\caption{\label{tab:table1}The best fitted values for the couplings
are listed here along with the chosen Borel window. Overal shift of the
Borel window by 0.5 GeV$^2$ to higher mass region gives 10 \% error.
 }
\begin{ruledtabular}
\begin{tabular}{ccc}
Borel Window (GeV$^2$) & fitted value \\
\hline
$2.5 \le M^2 \le 3.5$  & $f_D f_{D^*} g_{DD^*\pi}=0.31$ GeV$^2$\\
$3\le M^2 \le 4$  & $f_{D_1}f_{D^*}g_{D_1D^*\pi}=0.948$ GeV$^3$\\
$2\le M^2 \le 3$  & $f^2_{D^*}g_{D^*D^*\pi}=0.016$ GeV$^2$\\
$2\le M^2 \le 3$  & $f^2_{D_1}g_{D_1D_1\pi}=0.012$ GeV$^2$\\
\end{tabular}
\end{ruledtabular}
\end{table}

\begin{table}
\caption{\label{tab:table2}Comparison of our results with the other
published results. The results of
Refs.\cite{Belyaev:1995zk,Colangelo:1998rp,Aliev:1997bp}
are from light-cone QCD sum rules, the results from
Refs.\cite{Colangelo:1994es,DiBartolomeo:1995ir} are 
from the conventional sum rules, and the result of Ref.\cite{Navarra:2000ji}
is from the three-point sum rules. 
 }
\begin{ruledtabular}
\begin{tabular}{ccccc}
   & $g_{DD^*\pi}$ & $g_{D_1 D^* \pi}$ (GeV) & 
$g_{D^* D^* \pi}$ & $g_{D_1 D_1 \pi}$ \\
\hline
this work & $8.2\pm 0.1$ & $15.8 \pm 2$ & $0.3\pm 0.03$ & $0.17 \pm 0.04$\\
Ref.\cite{Belyaev:1995zk} & $12.5\pm 1$ & & & \\
Ref.\cite{Colangelo:1998rp} & $11.85\pm 2.1$ &   & $0.31\pm 0.08$ & \\
Ref.\cite{Aliev:1997bp} &  & $10\pm 2$   & & \\
Refs.\cite{Colangelo:1994es,DiBartolomeo:1995ir} & $9\pm 2$ & 
& $0.35\pm 0.08$ & \\
Ref.\cite{Navarra:2000ji} & $5.7\pm 4$ &   & & \\
\end{tabular}
\end{ruledtabular}
\end{table}

\begin{figure}
\caption{The Borel curves for the $D^*D^*\pi$ and $D_1D_1\pi$ couplings
given in  Eqs.~(\ref{dsdspisum}) (\ref{11pisum}).
Here the RHS of the equations are plotted with respect to
the Borel mass $M^2$.
}
\label{fig4}

\setlength{\textwidth}{6.1in}   
\setlength{\textheight}{9.in}  
\centerline{%
\vbox to 2.4in{\vss
   \hbox to 3.3in{\includegraphics{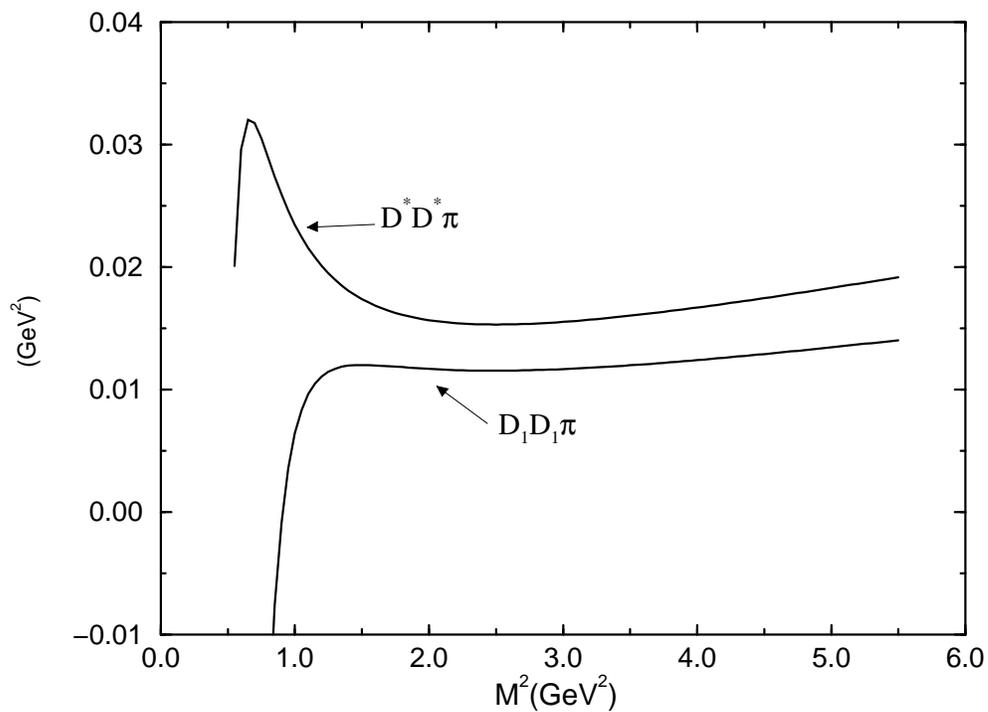}\hss}}
}

\vspace{100pt}
\end{figure}

\end{document}